\documentclass[aps,pra,superscriptaddress,twocolumn,amsmath,amssymb,floatfix]{revtex4}
\usepackage{graphicx}% Include figure files
\usepackage{dcolumn}% Align table columns on decimal point
\usepackage{bm}% bold math
\usepackage{subfig}
\usepackage{caption}
\usepackage{amsmath}
\usepackage{amssymb}
\usepackage{array}
\usepackage{ulem}
\usepackage{float}
\usepackage{ulem}
\newcommand{\ket}[1]{|{#1}\rangle}

\begin{document}
\title{Rydberg-Blockade Effects in Autler-Townes Spectra of Ultracold Strontium }
\author{B. J. DeSalvo}
\author{J. A. Aman}
\affiliation{Rice University, Department of Physics and Astronomy, Houston, Texas, USA 77251}
\affiliation{Rice University, Rice Center for Quantum Materials, Houston, Texas, USA 77251}
\author{C. Gaul}
\author{T. Pohl}
\affiliation{Max Planck Institute for Complex Systems, Dresden, Germany}
\author{S. Yoshida}
\author{J. Burgd\"orfer}
\affiliation{Institute for Theoretical Physics, Vienna University of Technology, Vienna, Austria, EU}
\author{K. R. A. Hazzard}
\affiliation{Rice University, Department of Physics and Astronomy, Houston, Texas, USA 77251}
\affiliation{Rice University, Rice Center for Quantum Materials, Houston, Texas, USA 77251}
\author{F. B. Dunning}
\affiliation{Rice University, Department of Physics and Astronomy, Houston, Texas, USA 77251}
\author{T. C. Killian}
\affiliation{Rice University, Department of Physics and Astronomy, Houston, Texas, USA 77251}
\affiliation{Rice University, Rice Center for Quantum Materials, Houston, Texas, USA 77251}
\date{\today}

\begin{abstract}
We present a combined experimental and theoretical study of the effects of Rydberg interactions on Autler-Townes spectra of ultracold gases of atomic strontium. Realizing two-photon Rydberg excitation via a long-lived triplet state allows us to probe the thus far unexplored regime where Rydberg state decay presents the dominant decoherence mechanism. The effects of Rydberg interactions are observed in shifts, asymmetries, and broadening of the measured atom-loss spectra. The experiment is analyzed within a one-body density matrix approach, accounting for interaction-induced level shifts and dephasing through nonlinear terms that approximately incorporate correlations due to the Rydberg blockade. This description yields good agreement with our experimental observations for short excitation times. For longer excitation times, the loss spectrum is altered qualitatively, suggesting additional dephasing mechanisms beyond the standard blockade mechanism based on pure van der Waals interactions.
\end{abstract}

\maketitle

\section{Introduction}
Long-range interactions between Rydberg atoms give rise to the Rydberg-blockade effect \cite{gmw09,ujh09,Comparat10}, which is of interest for quantum information \cite{swm10}, quantum optics (e.g. \cite{pfl12}), dynamics of driven dissipative systems \cite{app07,app07b,asp11,Lesanovsky11,lhc12,hon13,sgr14,mvs14,haf15,url15,lga14,sgw15}, and many-body physics with long-range interactions. The latter category includes transitions to ordered phases of Rydberg excitations or atoms \cite{pdl10,Bijnen11,sce12,sce15,Henkel10,Pupillo10}, realization of spin-models on optical lattices \cite{gdn15,bpo15}, and phenomena in gases such as three-dimensional solitons  \cite{Maucher11}, roton-maxon excitations \cite{Henkel10}, and super-solid states \cite{Henkel10,Pupillo10,Cinti10,BoninsegniRMP12}.  Controlling the strength and shape of  interactions by mixing a small amount of Rydberg character into atomic ground-state wavefunctions using off-resonant optical excitation (``Rydberg dressing" \cite{Henkel10,Honer10,Johnson10,Balewski14,Jau15}) figures prominently in most of these proposals.

In spite of recent advances \cite{Jau15}, the controlled generation of unitary interactions in large ensembles remains elusive  because of the large loss and dephasing rates observed experimentally \cite{wnm12,Balewski14}. Much remains to be understood, especially on how complex {processes} in dense Rydberg gases  affect these systems. These processes include plasma formation \cite{rtn00,kpp07,wnm12}, non-adiabatic level-crossings at short-range \cite{url15}, and superradiance \cite{wyc07,wpa08,crw13,Karlewski15}. The correct description of the correlations induced by Rydberg blockade and Rydberg dressing in very dense gases with strong Rydberg excitation is also an active area of study \cite{Johnson10,Balewski14}.

Many of these open questions have been studied via two-photon Rydberg excitation of alkali atoms in a three-level ladder configuration, where different regimes corresponding to coherent population trapping (CPT) \cite{sgh10,sap11}, electromagnetically induced transparency (EIT) \cite{wpa08,rhw09,pmg10,sap11}, and Autler-Townes (AT) spectroscopy \cite{agh10,zwc13,zzw14} can be accessed by varying the relative intensity of the two excitation lasers. For alkali atoms this typically involves a long-lived Rydberg state and a much more rapidly decaying intermediate state, which for example requires a strongly driven low lying transition in order to resolve the structure of AT spectra.

A comprehensive theoretical description of such interacting three-level systems in the presence of dissipation and strong correlations remains a challenge. Previous work studied different regimes and succeeded to describe certain aspects of experiments, e.g., through low-intensity expansions \cite{sev11}, classical Monte Carlo simulations \cite{app07,gar14,sch14}, density-matrix cluster expansions in the limit of low densities \cite{sgh10} or quantum trajectory Monte Carlo simulations of small systems \cite{hon13,gar14}. An insightful approach to analyze experimental observation is based on the corresponding single-body optical Bloch equations augmented by additional terms describing interaction induced Rydberg level shifts as well as dephasing of the Rydberg transition \cite{wpa08,rhw09,crw13,bhl13,zwc13,zzw14}. Depending on the particular setting such measurements where found to be consistent with either interaction induced line shifts \cite{crw13} or pure dephasing \cite{wpa08,rhw09,bhl13,zwc13,zzw14}.

While most experimental work has focussed on alkali Rydberg gases, alkaline-earth metal atoms have attracted significant interest recently because of new possibilities offered by their divalent electronic structure. The principal transition of the Rydberg core, which is typically in the visible, can be used to drive auto-ionizing transitions \cite{mlj10}, to image Rydberg atoms or ions \cite{mzs13}, and to provide oscillator strength for magic-wavelength optical trapping of Rydberg atoms \cite{mmn11}. Moreover, there is a greater variety of Rydberg-Rydberg interactions available because of the  existence of triplet and singlet excited levels \cite{vjp12}. Compared to alkali atoms \cite{hnp10}, two-photon excitation to triplet Rydberg levels via a long-lived triplet state, as demonstrated here, can also reduce the decoherence from light scattering for a given Rydberg state coupling and therefore holds promise for Rydberg dressing in such systems.

In this work we present the first experimental study of Rydberg-Rydberg atom interactions in cold alkali earth gases excited via long-lived triplet states. We probe such interactions via AT spectroscopy of dense gases through direct measurements of atomic loss. As a unique feature of our experiment, the long lifetime of the intermediate state enables AT spectroscopy for strongly driven Rydberg transitions, i.e., in the regime of small Rydberg state population similar to EIT experiments in alkali gases.
In contrast to previous studies of related alkali systems \cite{wpa08,rhw09,crw13,bhl13,zwc13,zzw14}, our measurements show clear signatures of both level shifts as well as decoherence induced by the strong Rydberg-Rydberg atom interactions. Our theoretical analysis is based on an effective one-body description augmented by nonlinear energy shifts and dephasing rates that are proportional to the Rydberg density and obtained from a meanfield description accounting for excitation blockade effects \cite{wlp08}. A comparison to the experimental results for short evolution times suggests that the nonlinear shift and dephasing rate are of equal magnitude and consistent with the calculated value of van der Waals interactions \cite{vjp12} and associated blockade radius. At long evolution times we observe an additional loss feature on two photon resonance that can not be explained by the sole action of van der Waals interactions between the laser-excited Rydberg states, and we provide possible explanations in terms of additional dephasing mechanisms. A deeper understanding and ultimately the control of the observed loss will be important, e.g., for future applications of Rydberg dressing in such systems.

\section{Experimental Methods}
\label{sec:expmethods}
 We perform our experiments on $^{84}$Sr atoms confined in an optical dipole trap (ODT) formed by crossed 1064-nm laser beams with waists of 300 $\mu$m (65 $\mu$m), and 440 $\mu$m (38 $\mu$m) in the horizontal (vertical) dimension.  These beams propagate in the horizontal plane and cross at a 90$^\circ$ angle.  A 1 s hold in the ODT results in several million atoms at a temperature of 700 nK. Evaporative cooling for 5 seconds  produces pure condensates of $4 \times 10^5$ atoms. For the experiments described here, however, we halt the evaporation before a BEC forms and conduct measurements on a sample with temperature $T\approx 150$\,nK. The details of cooling and trapping $^{84}$Sr are described elsewhere \cite{mmy09,desalvothesis}.

\begin{figure}[ht]
\centering
\includegraphics[width = 3 in]{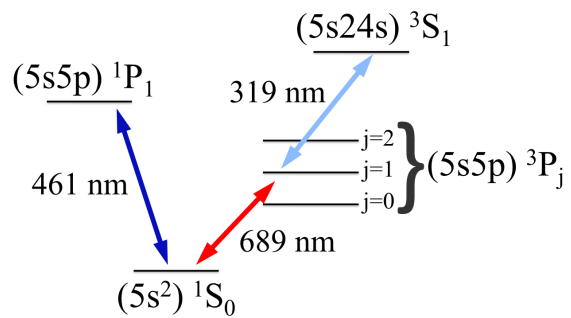}
\caption{Partial level diagram for Sr showing all transitions discussed in the text.}
\label{fig:LevelDiagram}
\end{figure}

We  excite atoms exclusively to the $5s24s\,^3\!S_1$ Rydberg state (lifetime $\tau_{^3\!S_1}\approx 4 \mu$s \cite{Kunze93}) with two-photon excitation using the narrow $^1\!S_0 \rightarrow {^3\!P_1}$ transition ($\tau_{^3\!P_1}=21\,\mu$s) as the intermediate state (Fig.\ \ref{fig:LevelDiagram}).
The first transition is driven using 689 nm light from the laser used for intercombination-line laser cooling. The Rabi frequency for this beam, $\Omega_{01}$, is
determined with an accuracy of 15\% by measuring the frequency of $^1\!S_0 \rightarrow {^3P_1}$ Rabi oscillations. For these experiments, we use values of $\Omega_{01}/2 \pi$ between  26 and 133 kHz and {the detuning of the 689-\,nm laser from resonance} ($\Delta_{01}/2\pi$) is between   -2 and +2 MHz.

\begin{figure}[ht]
\centering
\includegraphics[width = 3 in]{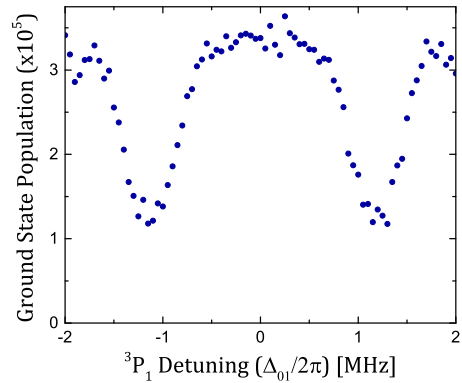}
\caption{Autler-Townes spectrum for weak excitation ($\Omega_{01}/2\pi=26$\,kHz) in a low density ($\rho=1.9\times 10^{12}$\,cm$^{-3}$) sample with the 319 nm laser on resonance with the $5s5p\,{^3\!P_1} \rightarrow 5s24s\,{^3\!S_1}$ transition ($\Delta_{12} = 0$).  The loss peaks are spaced by the UV Rabi frequency, $\Omega_{12}/2\pi=2.4$\,MHz. }
\label{fig:OnResonance}
\end{figure}

The upper leg of the two-photon transition is strongly driven using 319 nm radiation generated by frequency-doubling  light at 638 nm obtained by sum-frequency mixing of the pump and signal beams in an optical parametric oscillator pumped by a single-frequency fiber laser at 1064 nm.
%For frequency stabilization and tuning, we offset lock the laser to an optical cavity that is stabilized with the 689\,nm  light, and
The UV radiation has a full-width-half-maximum of 300\,kHz and is held on resonance  with detuning $\Delta_{12} = 0$.
The UV light has a beam waist of $600\,\mu$m at the atoms and a power of 34\,mW, resulting in a Rabi frequency $\Omega_{12} /2 \pi = 2.4$ MHz, as determined from the separation of the loss peaks in the AT spectrum (See Fig.\ \ref{fig:OnResonance}).
 The laser beam intensity profiles are much broader than the size of the trapped atom sample ($\sim 45\mu\mathrm{m} \times 30\mu\mathrm{m} \times 4 \mu\mathrm{m}$), so we neglect spatial variation of the Rabi frequencies.

We apply a magnetic field of 1.5\,G in the vertical direction, which defines our quantization axis.
% and shifts the frequency of the $^1\!S_0 \rightarrow ^3\!P_1 (m_j =1)$ transition approximately 3 MHz from its unperturbed frequency.
 Electric fields of up to 0.8 V/cm can be applied parallel to the magnetic field using field plates located outside the vacuum viewport windows.
The 689 nm light propagates anti-parallel to gravity and is  circularly polarized to drive the $^1\!S_0 \rightarrow {^3\!P_1} (m_j = + 1)$ transition.  The 319 nm light propagates horizontally and is vertically polarized  to drive the $^3\!P_1\, (m_j = +1) \rightarrow 5s24s\,^3\!S_1\, (m_j = + 1)$ transition.   The timing and power of both lasers are precisely controlled by acousto-optic modulators. The optical dipole trap is left on during excitation, and all detunings are measured with respect to line centers that include the AC Stark shift.

After excitation, the atoms are released from the trap, and the ground-state atom population is measured with time-of-flight absorption imaging on the $5s^2$\,$^1S_0$-$5s5p$\,$^1P_1$ transition at 461\,nm. Excitation to the Rydberg level is detected as ground-state atom loss, which can result  from direct trap loss through recoil, from decay to the very long lived $5s5p\,({^3\!P_0},{^3\!P_2})$ states, and through inelastic collisions \cite{dad15}.

\begin{figure}[ht]
\centering
\includegraphics[width = 3.2 in]{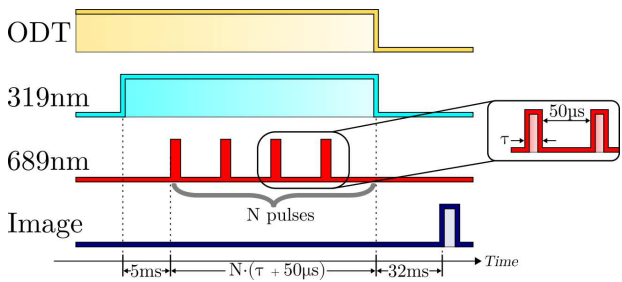}
\caption{Timing diagram for recording Autler-Townes spectra.  Details in the text. }
\label{fig:SpectraTiming}
\end{figure}

For recording loss spectra, we employ the pulsed excitation scheme shown in Fig.\ \ref{fig:SpectraTiming}.  We first turn on the 319 nm laser. 5 ms later, we apply a series of $N$ pulses of the 689 nm laser, with   $N$ chosen to yield approximately 50\% peak loss.  After the pulse sequence, the UV light remains on for $50\,\mu$s. All light is then extinguished, and the atoms are released from the trap and imaged after a 32 ms time of flight.

The  689\,nm pulses have a preselected ``on" time followed by 50 $\mu s$ of ``off" time between each pulse.  Since our method of detection is counting remaining ground state atoms, shot-to-shot atom-number fluctuations and other technical sources of noise make it hard  to detect the excitation of a small number of Rydberg atoms in a single pulse.
 {By using $N$ pulses, we amplify the loss to get a better signal-to-noise ratio. Because the off-time is chosen to be long compared to the lifetime of both the $^3\!P_1$ and $5s24s\,^3\!S_1$ states,  the conditions at the start of each pulse are identical up to a change of the total number of atoms. Thus to a good approximation, the series of short pulses only amplifies the signal in contrast to simply using a longer pulse, which would modify the physics by increasing the excitation fraction.}
To compare the data to theory at each data point, we assume an exponential decay of atom number with excitation time to estimate the fractional loss for a single pulse.

\begin{figure}[ht]
\centering
\includegraphics[width = 3.2 in]{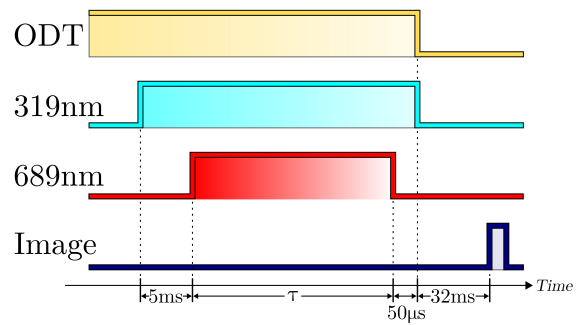}
\caption{Timing diagram for recording time evolution data.  Details in the text. }
\label{fig:TimeTiming}
\end{figure}

For taking time-evolution data, the pulse sequence in Fig.\ \ref{fig:TimeTiming} is employed. We first turn on the 319 nm light for 5 ms, and then apply a single pulse of 689 nm light.  This pulse is then followed by 50 $\mu s$ of just 319 nm light before both the UV and ODT beams are extinguished and the atoms are allowed to fall for a 32 ms time of flight before being imaged.

\section{Theoretical Description}

\label{sec:theory}

\subsection{Single-Particle Density-Matrix Treatment}

 \begin{figure}[ht]
\centering
\includegraphics[width = 1.75 in]{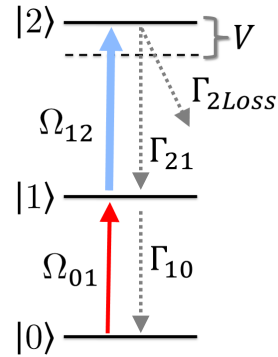}
\caption{Simplified level diagram of the three-level system used to model AT spectra. }
\label{fig:OpticalBlochTerms}
\end{figure}

 In a non-interacting or very low-density gas, the AT spectrum for the UV laser on resonance consists of  two symmetric loss peaks split by the UV Rabi frequency, $\Omega_{12}/2\pi$.  To model the effects of interactions, we calculate the evolution of the density matrix, $\sigma$, for a three-level system (Fig.\ \ref{fig:OpticalBlochTerms}) including
  {an approximate treatment of interactions between atoms in the Rydberg state, $\ket{2}$, including  shifts and a phenomenological dephasing term \cite{Tanasittikosol12,Karlewski15}. {This approximate treatment is formally similar to a mean-field theory, but includes a simple approximation to the two-body correlations, which are essential to reproduce the scalings found in experiment.} The details are discussed in Sec.\ \ref{sec:modifiedmeanfield}.}
 Non-unitary terms such as spontaneous decay and decoherence can be described using an appropriately chosen Lindblad superoperator, $\mathcal{L}(\sigma)$ in the master equation,
$\dot{\sigma} = \frac{i}{\hbar}[\sigma,H] + \mathcal{L}(\sigma)$.
For our system, this results in
 the optical Bloch equations \cite{Tanasittikosol12,Karlewski15},

% {(We can match notation to Christopher and Thomas' paper. }

\begin{flalign}
 \dot{\sigma}_{00}  = &  \Gamma_{10} \sigma_{11}  -  {\Omega}_{01} \operatorname{Im}(\sigma_{01}) & \nonumber \\
 \dot{\sigma}_{11}  = & -\Gamma_{10} \sigma_{11} + \Gamma_{21} \sigma_{22} + {\Omega}_{01} \operatorname{Im}(\sigma_{01}) - {\Omega}_{12} \operatorname{Im}(\sigma_{12}) & \nonumber \\
 \dot{\sigma}_{22}  = & -(\Gamma_{21} + \Gamma_{2 \mathrm{Loss}}) \sigma_{22} +  {\Omega}_{12} \operatorname{Im}(\sigma_{12}) & \nonumber \\
 \dot{\sigma}_{01}  = & -\Big(\frac{\Gamma_{10} + \Gamma_{689}}{2} + i \Delta_{01}\Big)\sigma_{01} - \frac{i {\Omega}_{01}}{2} (\sigma_{11} - \sigma_{00})& \nonumber\\
  &\hspace{0.2in}+ \frac{i {\Omega}_{12}}{2} \sigma_{02} & \nonumber \\
 \dot{\sigma}_{12}  = & - \Big[\frac{\Gamma_{10} +\Gamma_{21} + \Gamma_{2 \mathrm{Loss}} + \Gamma_{319} + \Gamma_{\mathrm{Ryd}}\sigma_{22}}{2} & \nonumber \\
 &\hspace{0.2in}+ i (\Delta_{12}-V_{\mathrm{Ryd}} \sigma_{22})\Big]\sigma_{12} - \frac{i {\Omega}_{12}}{2}(\sigma_{22} - \sigma_{11}) & \nonumber \\
  &\hspace{0.2in}  - \frac{i {\Omega}_{01}}{2} \sigma_{02} & \nonumber \\
 \dot{\sigma}_{02}   = & -\Big[\frac{\Gamma_{689} + \Gamma_{21} + \Gamma_{2 \mathrm{Loss}} + \Gamma_{319} + \Gamma_{\mathrm{Ryd}} \sigma_{22}}{2} & \nonumber \\
 &\hspace{0.2in}+ i(\Delta_{01} +\Delta_{12} - V_{\mathrm{Ryd}}\sigma_{22})\Big]\sigma_{02} + \frac{i {\Omega}_{12}}{2}\sigma_{01} & \nonumber \\
&\hspace{0.2in} - \frac{i {\Omega}_{01}}{2}\sigma_{12}&
\label{eq:OpticalBloch}
\end{flalign}
where $\Gamma_{ij}$ denotes the spontaneous decay rate from $\ket{i} \rightarrow \ket{j}$, $\Gamma_{2Loss}$ denotes spontaneous decay from $\ket{2}$ that results in a loss of atoms such as decay into $5s5p\,{^3P_2}$ and $5s5p\,{^3P_0}$ states, and $\Gamma_{689}$ and $\Gamma_{319}$ are the dephasing rates due to respective laser linewidths.  The Rabi frequencies  are time-dependent to accommodate non-simultaneous laser on and off times.

From fitting our low density and weak excitation date, we estimate
$\Gamma_{689}= 120 \times 10^3 s^{-1}$ and $\Gamma_{319}= 1.2 \times 10^6 s^{-1}$. This is consistent with
 analysis of our laser lock circuitry and other spectral measurements.
The decay rate of the $^3\!P_1$ state is known very accurately.  However,  the decay rate of the excited state is less well known. We obtain the best agreement with our data  using a decay rate of $(\Gamma_{21} + \Gamma_{2 Loss})= 310 \times 10^3 s^{-1}$, which is slightly higher than the natural decay rate expected from scaling the results of  \cite{Kunze93}. The  branching ratio calculated from the Wigner-Eckart theorem would imply that 1/3 of the decays from the Rydberg state result in $^3P_1$ atoms, but the recoil energy for a single 320\,nm photon exceeds the trap depth, so we assume all radiative decay leads to atom loss and set $\Gamma_{21}=0$. This slightly improves agreement between simulation and data.
%We use a branching ratio of $\Gamma_{2 Loss}/\Gamma_{21}=2$ calculated using the Wigner-Eckart theorem assuming loss results from decay  to $5s5p\,({^3P_0},{^3P_2})$ levels.

In Eq.\ \ref{eq:OpticalBloch}, $V_{Ryd} \sigma_{22}$ describes the level shift due to interactions (in units of radial frequency) within our approximations, and $ \Gamma_{Ryd} \sigma_{22}$ describes a phenomenological dephasing due to Rydberg-Rydberg interactions.
The influence of each will depend on the density of Rydberg atoms through  the factor $\sigma_{22}$ and whatever density dependence is in $V_{Ryd}$ and $\Gamma_{Ryd}$, but to understand the effects of these terms  we first display AT spectra calculated with  $V_{Ryd} \sigma_{22} $ and $ \Gamma_{Ryd} \sigma_{22}$ replaced with constants
V and $\Gamma$.
\begin{figure}[ht]
\centering
\includegraphics[width = 3 in]{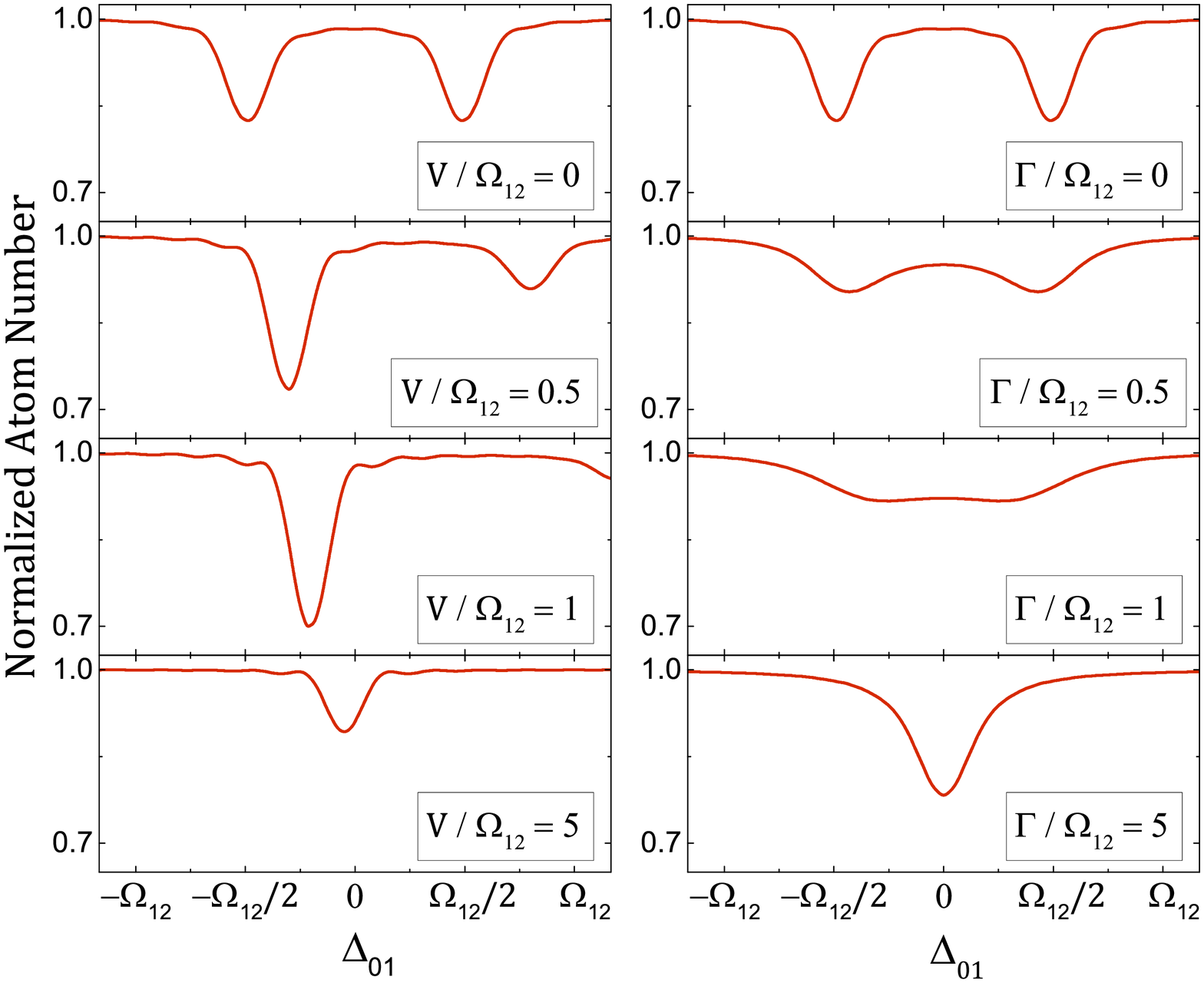}
\caption{Calculated effects of a constant level shift ($V$) (Left) or dephasing ($\Gamma$) (Right) on Autler-Townes  spectra for $\Delta_{12}=0$, a single excitation pulse of duration $t=2\,\mu$s, and Rabi frequencies $\Omega_{01}/2\pi=133$\,kHz and $\Omega_{12}/2\pi=2.4$\,MHz.  The values of $V$ and $\Gamma$ are as indicated.}
\label{fig:IntAndDephase}
\end{figure}
In the left panel of Fig.\ \ref{fig:IntAndDephase}, we see
 that increasing $V$ leads to a blue shift and an asymmetry in the spectrum,  similar to increasing the detuning of the UV laser, $\Delta_{12}$.  At large $V$, loss occurs close to two-photon resonance ($\Delta_{01}\approx0$), but it is strongly suppressed.  In our system, loss only occurs from the Rydberg state, and for large
$\Delta_{12}$ or $V$, the eigenstate resonantly excited at $\Delta_{01}\approx0$  has diminishing  Rydberg character.

The effect of increased dephasing is markedly different.  There is no shift and no asymmetry in the loss features, and at low values of $\Gamma$ only a reduction of the splitting and a broadening of the peaks are evident.  For large values of $\Gamma$, a very strong loss feature arises at two-photon resonance implying a large fraction of Rydberg character in the excited state. This can be understood in the limit of extremely large dephasing, in which the coherence between the intermediate and Rydberg states is never formed.
 Excitation to the intermediate state results in loss without the appearance of AT splitting.  From these plots, it is evident that the effects of $V$ and $\Gamma$ are largely separable in AT spectra.

%\subsection{Rydberg-Rydberg Interactions}
%To go forward with our analysis, we need to consider the functional form of Rydberg-Rydberg interactions.  A detailed treatment of this problem can be found in \cite{Comparat10}, but for our purposes it is sufficient to discuss the highlights.  In the limit of a large separation ($\vec{R} = R\vec{n}$) between two atoms, the interaction term of the Hamiltonian can be written as
%\begin{equation}
%H_{12} = \frac{\vec{\mu_1} \cdot \vec{\mu_2} - 3(\vec{\mu_1}\cdot \vec{n})(\vec{\mu_2}\cdot \vec{n})}{4\pi\epsilon_0 R^3}
%\label{eq:InteractionHamiltonian}
%\end{equation}
%where $\mu_i$ the dipole moment of atom $i$.
%
%In the atomic pair state basis, $\bra{r_1,r_2}$ where r labels the Rydberg state, we can rewrite the Hamiltonian in the convenient form
%\begin{equation}
%H_{int} =
%\Bigg(\begin{matrix}
%\hbar\Delta & V_{dip} \\
%V_{dip}^\dagger & 0
%\end{matrix}\Bigg)
%\end{equation}
%where we have defined $V_{dip} = \bra{r_1',r_2'}H_{12}\ket{r_1,r_2}$ and

\subsection{Theory of Rydberg-Rydberg Interactions with Two-body Correlations Arising from Blockade}
\label{sec:modifiedmeanfield}
The interaction between two $5s24s{^3S_1}$  Rydberg atoms at an inter-atomic separation $r$ can be described with an isotropic,
repulsive  van der Waals interaction (in frequency units), $V(r) = {C_6}/{\hbar} {r^6}$, with %$C_6/h=1.4\,\mathrm{MHz}\,\mu\mathrm{m}^6$ \cite{vjp12}.  {(Should we give value in rad/s?)}
$C_6/\hbar=8.8 \times 10^6\,\mu\mathrm{m}^6 \,\mathrm{s^{-1}}$ \cite{vjp12}.
We begin by recalling the   mean-field theory, in which the many-body interacting system is replaced by a model of one atom in an {external potential determined by the average density of other particles},
\begin{equation}
V_{\mathrm{MF}} = \int d\mathbf{r}' V(\mathbf{r}')\langle n_2(\mathbf{r}') \rangle.
\end{equation}
For  a translationally invariant system,
 the  density of Rydberg atoms is taken as a constant $\langle n_2(\mathbf{r}')\rangle=\sigma_{22}\rho$.
The mean-field approximation neglects all correlations, which for the present system leads to severe inaccuracies. In particular, one finds that $V_{\text{MF}}$ diverges.
%\begin{align}\label{eq:naivemeanfield}
%V_{eff} &= \int_0^\infty d \mathbf{r} \frac{C_6}{\hbar r^6} \langle n_2(\mathbf{r}) \rangle & \nonumber \\
%& = \frac{C_6 \sigma_{22} \rho }{\hbar} 4 \pi \int_ {0}^\infty dr r^{-4} = \infty.
%\end{align}

This divergence results from a failure to reasonably describe short-range correlations.  We expect strong effects from the Rydberg blockade, which prevents excitation of a second Rydberg atom within a blockade radius $R_B = (C_6/2 \hbar \Omega_{12})^{1/6}$ of an atom that is already in the Rydberg state.  This creates spatial correlations in the positions of excited atoms, which are neglected in a mean-field treatment. We can approximately incorporate correlations into the description by introducing a short-range cutoff  to the spatial integral at $R_B$,
\begin{eqnarray}\label{eq:scaling}
V_{\mathrm{eff}} &=& \int_{R_B}^\infty d \mathbf{r} \frac{C_6}{\hbar r^6} \langle n_2(\mathbf{r}) \rangle = \frac{C_6 \sigma_{22} \rho}{\hbar} 4 \pi \int_ {R_B}^\infty dr r^{-4}  \\ \nonumber
& =& \frac{4\pi C_6}{3 \hbar R_B^3} \sigma_{22} \rho.
\end{eqnarray}
%\begin{equation}
%V_{eff} &= \int_{R_B}^\infty d \mathbf{r} \frac{C_6}{\hbar r^6} \langle n_2(\mathbf{r}) \rangle = \frac{C_6 \sigma_{22} \rho}{\hbar} 4 \pi \int_ {R_B}^\infty dr r^{-4} & \\ \nonumber
%& = \frac{4\pi C_6}{3 R_B^3} \sigma_{22} \rho.
%\label{eq:scaling}
%\end{equation}
%This is equivalent to the introduction of a pair-correlation function described in \cite{wlp08}.
This  goes beyond mean-field theory by incorporating hard-core pair correlations between the Rydberg states in a manner analogous to Ref.\ \cite{wlp08}. However, the approximations made are quite drastic: they incorporate only pairwise correlations, correlations are only between the Rydberg levels, and the correlations are imposed with a rather crude hard core step function. Remarkably, we will see that this approximate description of the Rydberg level shift due to interactions suffices to quantitatively reproduce much of our data.
 It yields $V_{Ryd} =  {4\pi C_6\rho}/{3 \hbar R_B^3} $ in Eq.\ \ref{eq:OpticalBloch}.

\begin{figure}[ht] %look in C:\Users\killian\Documents\EXPERIMENT\RYDBERGEXPERIMENT\EXPERIMENTALPLANS
\centering
\includegraphics[width = 3.25 in]{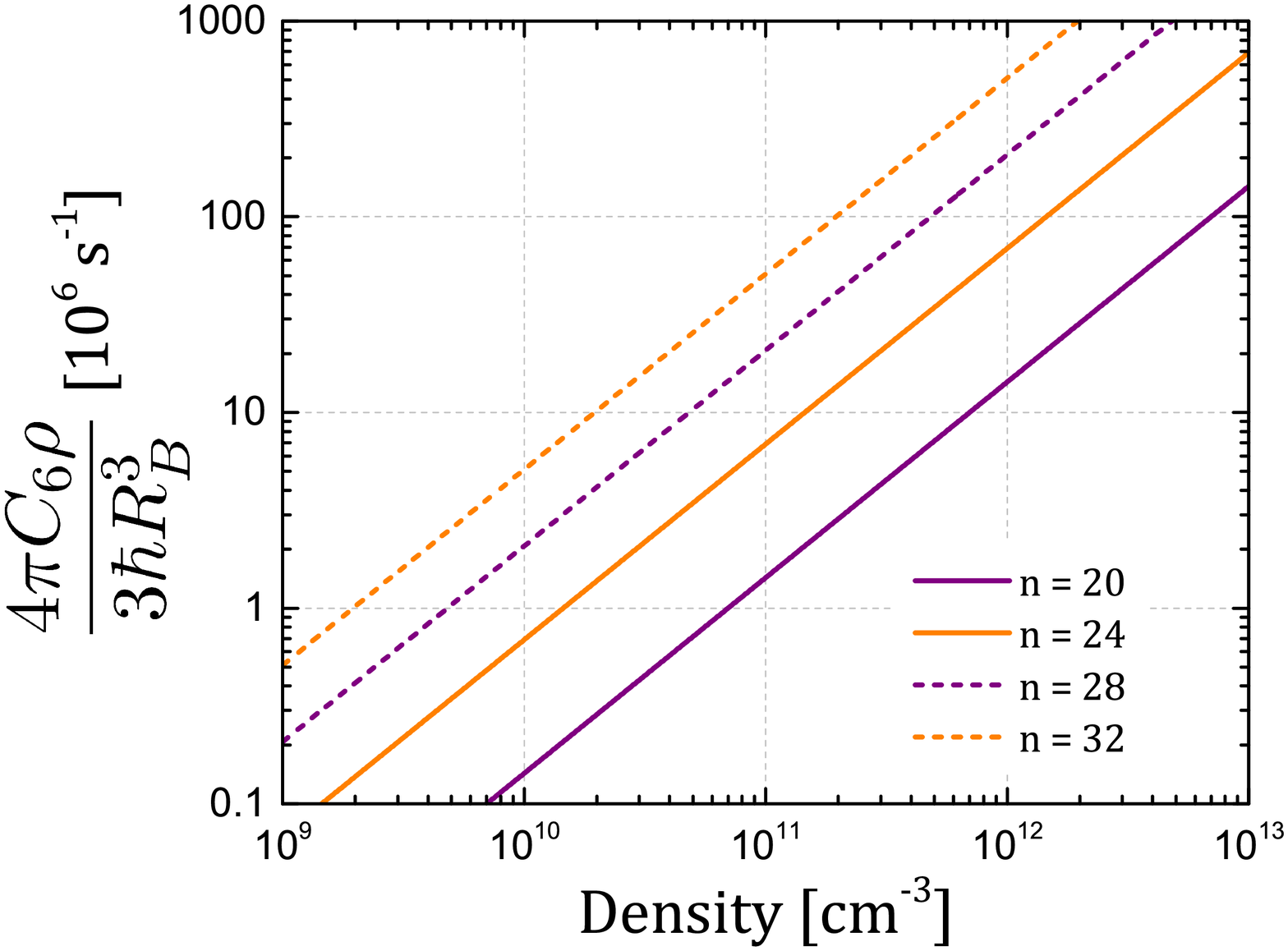}
\caption{Interaction energy coefficient (${4\pi C_6\rho}/{3 \hbar R_B^3 }$) predicted from the blockade-corrected mean field approximation in Eq.\ \ref{eq:scaling} as a function of  density ($\rho$) for $^3S_1$ Rydberg atoms with the indicated principal quantum numbers $n$ and $\Omega_{12}/2\pi=2.4$\,MHz.}
\label{fig:InteractionEnergy}
\end{figure}

 Figure \ref{fig:InteractionEnergy} shows the typical strength of interactions for various experimental conditions and the predicted $C_6$ value \cite{vjp12}. In our experiments, $R_B = 0.8\,\mu$m.
In this approach,
we have neglected effects of level crossings with other molecular potentials \cite{ufl15}, as well as effects due to higher order terms in the multipole expansion. Both   should only be important at internuclear distances less than $R_B$ for our experimental parameters.

{A unique feature of this work with strontium is that the linewidth of the intermediate $^3P_1$ state, $\Gamma_{10}=\Gamma_{^3P_1}=47\times 10^3$\,s$^{-1}$,  is much smaller than the linewidth of intermediate states used in experiments with alkali atoms. This allows us to be in the AT regime rather than the EIT regime even though we are strongly driving the intermediate-Rydberg transition. If one defines a general blockade radius as $R_B = (C_6/2 \hbar \gamma)^{1/6}$, then $\gamma=\Omega_{12}^2/\Gamma_{10}$ in the EIT regime, while $\gamma=\Omega_{12}$ in our experiments.}

 %the second is related to the choice of the blockade radius. There are several options being used in the literature (and some confusion attached to this) for the energy scale that defines R_b. In an optics context people use the EIT linewidth, Omega_2^2/gamma; for strong excitation people often use the two-photon coupling, Omega_1Omega_2/gamma; or just the upper Rabi frequency Omega_2 as you assume in the draft. In the present case, the correct choice nicely follows from the cluster-expansion and connects to other choices. As one sees even analytically, R_b is defined by the EIT width Omega_2^2/gamma as long as Omega_2<gamma but the energy scale crosses over to Omega_2 once Omega_2/gamma becomes greater than unity. For Sr the decay is very small such that the Rabi frequency is much larger which explains why the chosen blockade radius gives good results. I suppose if you take Omega^2/gamma as a cutoff energy scale the agreement would be much worse. For alkalines on the other hand one typically has Omega/gamma<1 such that the blockade radius will be different. I think describing this and presenting experimental evidence for this behaviour could make for a nice distinguishing feature of Alkaline-Earth gases.

Starting from this calculation of the level shift including correlation effects, we treat density inhomogeneity in our trapped gas within a local density approximation (LDA). The observable {$I$ (typically the total number of ground-state atoms remaining after a period of excitation in a trap) is calculated theoretically as an integral over density of $I$ evaluated} for fixed density, $I(\rho)$, with weighting determined by the distribution of densities in the trap,
\begin{equation}
I = \int_0^{\rho_{0}} \, I(\rho) g(\rho) \,d\rho,
\label{LDA}
\end{equation}
where  $\rho_0$ is the peak density and $g(\rho)$ is the weighting function
\begin{equation}
g(\rho) = \frac{2 \pi}{\omega_1\omega_2\omega_3}  \Big( \frac{2 k_B T}{m} \Big)^{3/2} [\ln(\frac{\rho_0}{\rho})]^{1/2},
\end{equation}
for  the  harmonic trap oscillation frequencies $\omega_{i}$.
 The length scale for density variation in the trap (given by the density distribution) is long compared to  the blockade radius, interatomic spacing, and the distance a Rydberg atom travels in its lifetime, justifying the use of the LDA. We numerically evaluate the integral in Eq.\ \ref{LDA} with an 11-point trapezoidal rule approximation.

While the form of $V_{Ryd}$ follows from a  physical model of the effects of Rydberg blockade on excitation correlations, a microscopic picture for
interaction-induced dephasing is less obvious. We qualitatively discuss possible sources of this dephasing in Sec.\ \ref{sec:resultsanddiscussion}.
Experimentally, we find that  most of our data can be well described assuming the dephasing is similar in magnitude to the level shift. We thus use $\Gamma_{Ryd} = \beta {4\pi C_6 \rho}/{3 \hbar R_B^3}$ in Eq.\ \ref{eq:OpticalBloch} and treat $\beta$ as an adjustable parameter.

%\subsection{Calculation Details}
%To simulate experimental results, the optical Bloch equations (Eq.\ \ref{eq:OpticalBloch}) are solved in Matlab to solve the equations for a set of
% 10 different densities, $\rho_i$, that are chosen such that
%\begin{equation}
%\int_{\rho_i}^{\rho_{i+1}} g(\rho) d\rho = N/10
%\end{equation}
%where N is the total number of particles.  We use these densities to perform a 10 point trapezoidal-rule approximation of the integral in equation \ref{LDA}.

%The following section will convince you of the necessity of including this term even though we don't understand its origin, and from the spectra you will see it is clear that the magnitude of the dephasing is at least on the order of the magnitude of $V$.  For that reason, we choose to treat $\Gamma$ in the same manner as $V$, that is with a linear scaling with density and within the LDA, yielding $\Gamma_{Ryd} = \beta \frac{4\pi C_6}{3 R_B^3} \rho$ in equation \ref{eq:OpticalBloch}.

\section{Results and Discussion}
\label{sec:resultsanddiscussion}

\subsection{Spectra for Short Excitation Time}
\begin{figure}[ht]
\centering
\includegraphics[width = 3 in, trim= 30 0 0 0]{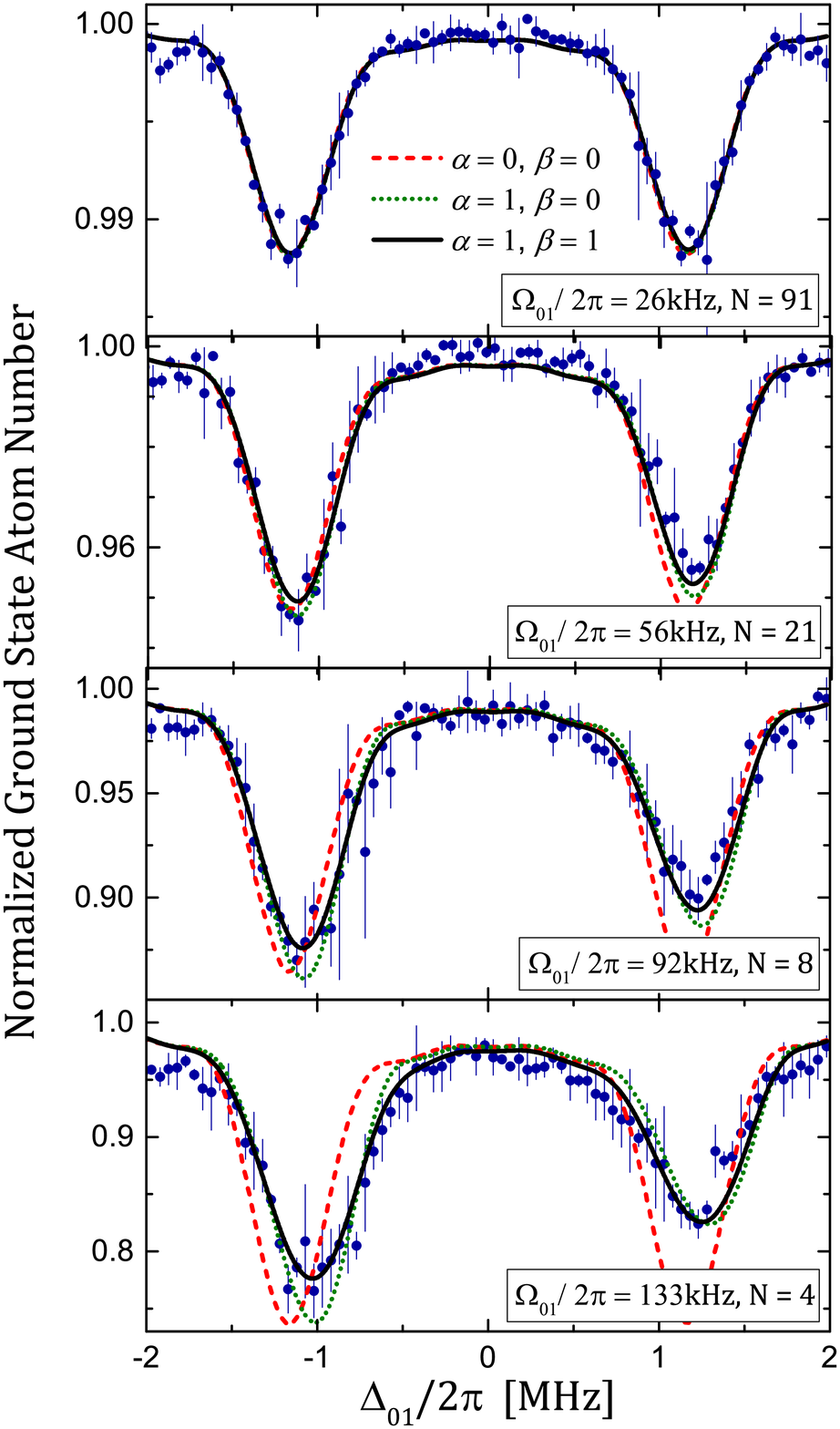}
\caption{Blue circles: Fractional number of ground state atoms remaining after  a single 2\,$\mu$s excitation pulse  and an initial peak density $\rho_0 = 1.9 \times 10^{12}$ cm$^{-3}$.   The  lines show the results of Eq.\ \ref{eq:OpticalBloch} and the LDA approximation for the interaction terms
$V_{Ryd} =  {\alpha 4\pi C_6\rho}/{3 \hbar R_B^3} $  and $\Gamma_{Ryd} =  {\beta 4\pi C_6\rho}/{3 \hbar R_B^3}$, with $\alpha$ and $\beta$ given in the legend.  {The simulation is performed for a single 2\,$\mu$s pulse of 689\,nm + 319\,nm excitation, followed by $50\,\mu$s of only UV light. The data represents the results of $N$ pulses, but the approximate fraction of atoms remaining after a single pulse is  plotted and calculated assuming the atom number at each frequency point decays exponentially in time.  $N$ is indicated in each plot. }}
\label{fig:LDShortSpectra}
\end{figure}

\begin{figure}[ht]
\centering
\includegraphics[width = 3 in, trim= 30 0 0 0]{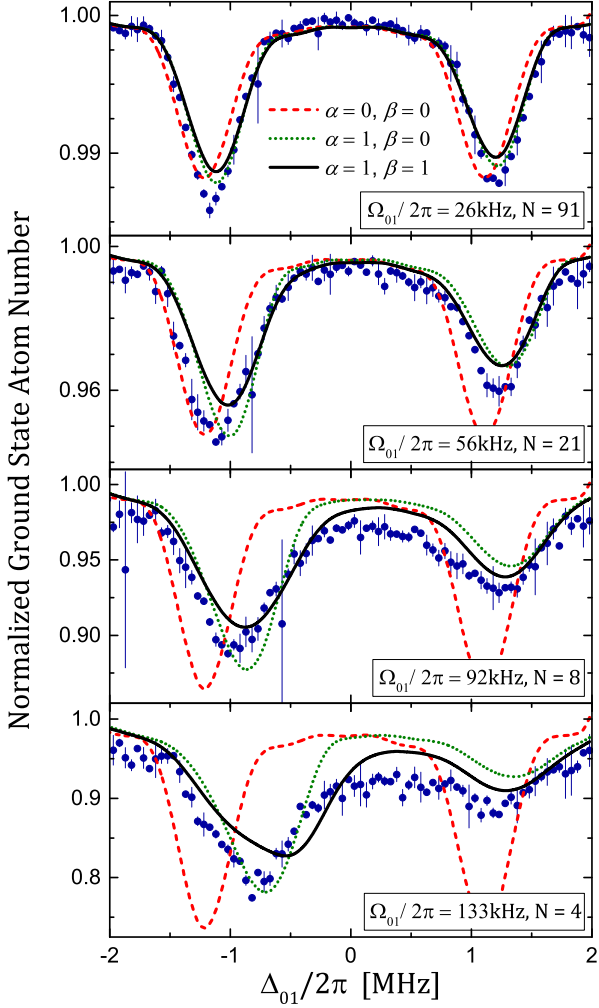}
\caption{Blue circles: Fractional number of ground state atoms remaining after a single 2\,$\mu$s excitation pulse and an initial peak density $\rho_0 = 1 \times 10^{13}$ cm$^{-3}$.     Theory lines show the results of Eq.\ \ref{eq:OpticalBloch} and the LDA approximation for the interaction terms $V_{Ryd} =  {\alpha 4\pi C_6\rho}/{3 \hbar R_B^3} $  and $\Gamma_{Ryd} =  {\beta 4\pi C_6\rho}/{3 \hbar R_B^3}$, with $\alpha$ and $\beta$ given in the legend. Treatment of the pulse sequences for simulation and data is the same as in Fig.\ \ref{fig:LDShortSpectra}.  }
\label{fig:HDShortSpectra}
\end{figure}

Short-time dynamics were probed using  excitation pulses of 2 $\mu s$ duration and the timing sequence illustrated in Fig.\ \ref{fig:SpectraTiming} and described in Sec.\ \ref{sec:expmethods}.  The UV laser is held on resonance $\Delta_{12}=0$ and the frequency of the 689\,nm laser is scanned to obtain AT spectra.
Spectra were recorded for a series of 689-nm laser intensities corresponding to $\Omega_{01}/2\pi=\{26,56,92,133\}$\,kHz.
Higher $\Omega_{01}$ in general corresponds to a higher Rydberg excitation rate.
{The number of pulses is adjusted for each spectrum to produce peak depletion of the ground state of about 50\% at the end of the pulse sequence.  However, to facilitate discussion and comparison with simulation, the loss after a single pulse is estimated assuming an exponential decay of atom number at each frequency point.}
%total loss is divided by the number of pulses so that the data are normalized to approximately show the fraction of atoms that remain after a single excitation pulse.
 Spectra are presented for two different peak densities, $\rho_0=1.9 \times 10^{12}$ cm$^{-3}$ (Fig.\ \ref{fig:LDShortSpectra}) and $\rho_0=1 \times 10^{13}$ cm$^{-3}$ (Fig.\ \ref{fig:HDShortSpectra}).
For principal quantum number $n = 24$, with $\Omega_{12} = 2.4$ MHz, the blockade density is $\rho_B = (4\pi R_B^3/3)^{-1} =4.2 \times 10^{11}$ cm$^{-3}$.  Therefore, for the presented peak densities the numbers of atoms within a blockade volume are ${\rho_0}/{\rho_B} \sim 5$ and $25$.  We note that the low and high density data are both recorded using the same  ODT configuration.
% and the variation in density was achieved by changing the length of the initial collection phase in the magnetic trap.  This results in a change in number and a slight change in temperature which gives rise to the two densities mentioned above.

For the low-density data (Fig.\ \ref{fig:LDShortSpectra}), the lowest excitation strength (top) results in an almost symmetric spectrum typical of a non-interacting gas.  Each loss peak at $\Delta_{01}=\pm \Omega_{12}/2$ represents excitation by the 689-nm laser to one of the dressed states $|^3P_1\rangle \pm |^3S_1\rangle$. The $2\,\mu$s excitation time is comparable to the lifetime of the Rydberg state, so the fraction of atoms lost is close to the relative population of the Rydberg level at the end of the pulse, which peaks at $\sim 10^{-2}$. This is supported by the numerical solutions of the optical Bloch equations. At the center of the spectrum ($\Delta_{01}=0$), vanishingly small loss rates $\sim f\Gamma_{2Loss}$ are expected from scattering in the wings of the symmetric peaks \cite{ads11}, where $f=\Omega_{01}^2/\Omega_{02}^2\sim 1\times 10^{-4}$ characterizes the fraction of population in the Rydberg state. The data are well described by the non-interacting single-particle density matrix treatment (Eq.\ \ref{eq:OpticalBloch}).
%, gives good confidence in the parameters that used in the model.
The density of Rydberg atoms and the excitation time are small enough that adding interaction terms ($V_{Ryd}={\alpha 4\pi C_6 \rho}/{3 \hbar R_B^3}$ and $\Gamma_{Ryd}=\beta{4\pi C_6 \rho}/{3 \hbar R_B^3}$ with $\alpha=\beta=1$) to the optical Bloch equations  has no significant effect.
We note that all simulations use  a value of $\Omega_{01}$ 16\% higher than determined from independent spectral measurements to get best agreement with the overall intensity of the signal. This is at the upper limit of our uncertainty, but is in reasonable agreement given the simplicity of our theoretical model.

%\{$V_{Ryd}=\frac{4\pi C_6 \rho}{3 \hbar R_B^3}$, $\Gamma_{Ryd}=0$\

With increasing $\Omega_{01}$, the spectra display a sizable shift and an asymmetry appears in the peak heights, which is a clear indication of  effects due to Rydberg-level shifts induced by Rydberg-Rydberg interactions as described in Fig.\ \ref{fig:IntAndDephase}. These effects are captured very well by adding the blockade-augmented mean-field interaction term $V_{Ryd}$ to the optical Bloch equations.  {The AT peaks display a pronounced shift to the blue. For a laser detuned slightly to the blue of each unperturbed resonance, this can be interpreted as an antiblockade effect as previously seen in an ultracold Rydberg gas \cite{app07,agh10}, in which the interactions shift levels into resonance with the laser.}

Adding interaction-induced dephasing through $\Gamma_{Ryd}$ with $\beta=1$ makes a small, but noticeable improvement by reproducing some of the broadening of the lines. At line center for the highest intensity excitation, the relative population of the Rydberg state is still small, $f=3\times 10^{-3}$,  and the probability of finding a second excited atom within  a blockade radius of a Rydberg atom for peak density  $f\rho_0/\rho_{blockade}$ is $\sim 1\times 10^{-2}$. So we do not expect Rydberg blockade effects to be important. On resonance with the two AT peaks, however, taking the Rydberg fraction as approximately equal to the loss fraction, we find $f\rho_0/\rho_{blockade}$ is on the order of one. This is consistent with the  strong interaction effects that are observed.

For the high-density sample (Fig.\ \ref{fig:HDShortSpectra}), the spectra show  shifts and asymmetries even for the lowest values of $\Omega_{01}$. These spectra are described reasonably well by including the blockade-augmented mean-field interaction term $V_{Ryd}$, just as in the low-density case. This confirms the linear scaling with density of the blockade-augmented mean-field level shift, which implies that our modified calculation of the interaction strength incorporating spatial correlations through a short-range cutoff on Rydberg-Rydberg distances (Eq.\ \ref{eq:scaling}) captures important aspects of the physics. We conclude that correlations due to the Rydberg blockade effect are playing a strong role during excitation on resonance with the AT peaks in this regime.
%{Our treatment of correlations (Eq.\ \ref{eq:scaling}) is arguably the simplest extension one could make beyond mean-field. So it is surprising that it works so well.}

The high density data also display clear signatures of dephasing through the broadening of the lines and increased atom loss at line center.   The inclusion of $\Gamma_{Ryd}$ with $\beta=1$ improves the agreement,  even for very strong interactions ($\Omega_{01}/2\pi=92$ and 133\,kHz). For this level of modification of the spectrum, the dephasing rate $\Gamma_{Ryd}$ must be interpreted as a phenomenological parameter. A microscopic description of such dephasing terms can be obtained from a more detailed calculation of two-body correlations \cite{sap11}, which goes beyond the scope of the present study.
%One possible microscopic origin for the broader spectra could be the variation of interactions due to the distribution of nearest-neighbor separations in even a homogeneous bulk gas. This variation can be on the order of the interaction strength itself \cite{desalvothesis}.
%One also expects decoherence to arise from many-body correlation effects, which are predicted by more advanced multi-particle theory \cite{sap11}.

% POTENTIAL FLUCTUATIONS DO NOT CAUSE DEPHASING. ON THE MF LEVEL THIS CAN BE SEEN DIRECTLY FROM A MICROSCOPIC (i-resolved) SIMULATION OF THE MF EQUATIONS WITH FLUCTUATING MF-POTENTIALS

There is a noticeable discrepancy near the center of the spectrum ($\Delta_{01}=0$), however, where the experimental data show much more loss than predicted by the simulation, suggesting stronger dephasing rates. To explore this effect, we recorded data with longer excitation times.

\begin{figure}[t]
\centering
\includegraphics[width = 3 in, trim=30 0 0 0]{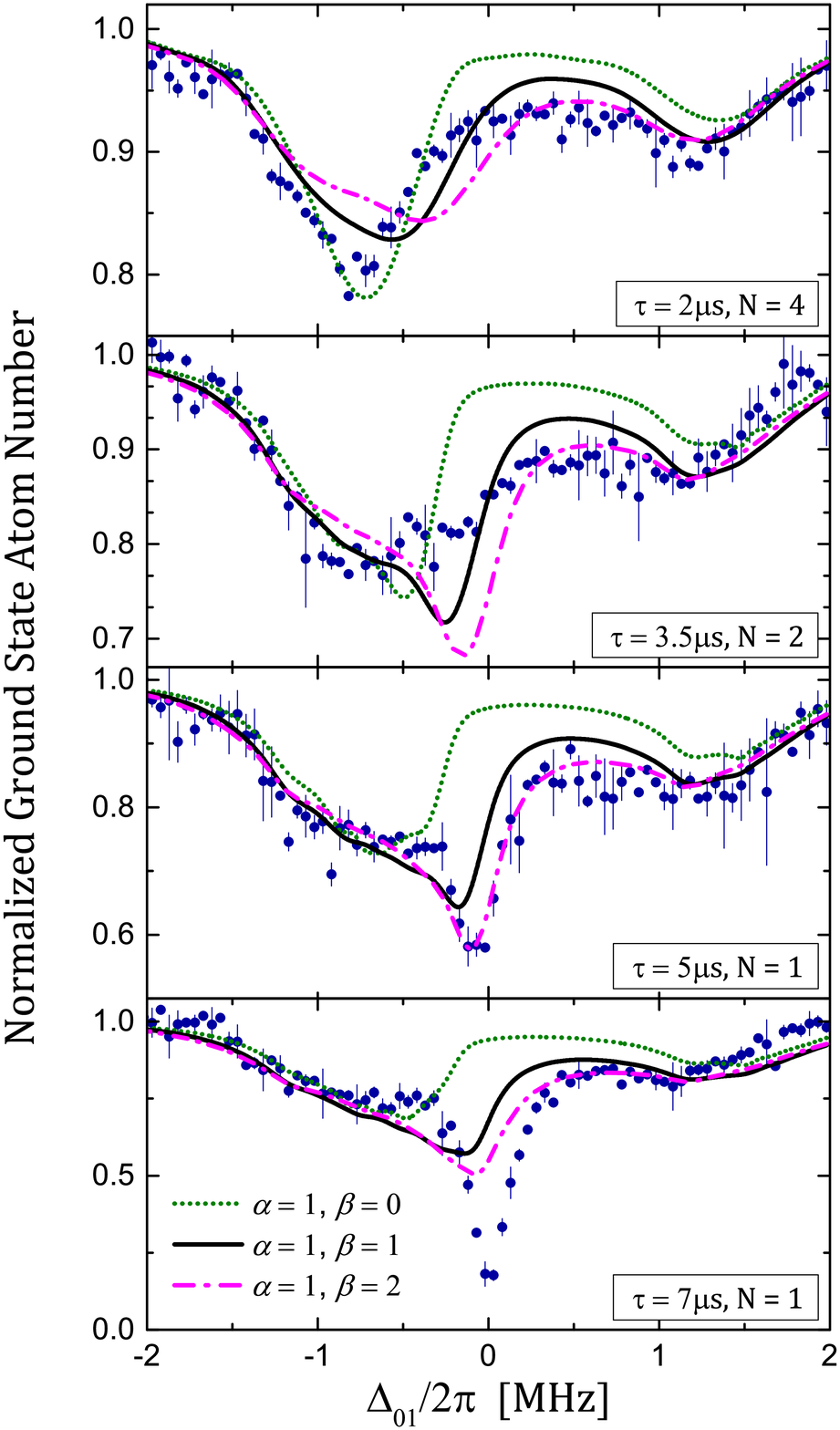}
\caption{Blue circles: Fractional number of ground state atoms remaining after a single  pulse of length indicated in each plot. The initial peak density is $\rho_0 = 1 \times 10^{13}$ cm$^{-3}$ and $\Omega_{01} = 133$\,kHz.  (More details are provided in the text.)  The lines show the results of Eq.\ \ref{eq:OpticalBloch} and the LDA approximation for the interaction terms
$V_{Ryd} =  {\alpha 4\pi C_6\rho}/{3 \hbar R_B^3} $  and $\Gamma_{Ryd} =  {\beta 4\pi C_6\rho}/{3 \hbar R_B^3}$, with $\alpha$ and $\beta$ given in the legend.
Treatment of the pulse sequences for simulation and data is the same as in Fig.\ \ref{fig:LDShortSpectra}.
 }
\label{fig:HDTimeSpectra}
\end{figure}

%Pulse Width [us]	Num. Pulses
%2	4
%3.5	2
%5	1
%7	1

\subsection{Spectra for Longer Excitation Times}
The short excitation time allows us to observe strong Rydberg interactions, but in a regime in which level shifts are still the dominant effect. At longer excitation times, however, dephasing dramatically alters the excitation dynamics.
To probe dynamics on a longer time scale, we obtained AT spectra at high peak density ($1 \times 10^{13}$ cm$^{-3}$) and high 689-nm intensity ($\Omega_{01}/2\pi=133$\,kHz) for a series of increasing excitation-pulse durations ranging from 2 to 7 $\mu$s  (Fig.\ \ref{fig:HDTimeSpectra}). Multiple pulses are applied, following the timing sequence described in Fig.\ \ref{fig:SpectraTiming}. As before, the number of pulses is adjusted for each spectrum to produce peak atom loss of about 50\% at the end of
the pulse sequence, and for display in the figure and comparison to theory,  the spectra are  normalized to show the fraction of atoms remaining after a single excitation pulse assuming exponential decay of atom number.

Dephasing corresponding to $\beta=1$ matches the data for 2\,$\mu$s excitation time   for detuning well removed  from the center of the spectrum, while increased dephasing ($\beta=2$) is required to reproduce the loss near $\Delta_{01}=0$. % {It is important to note that this level of modification of the spectrum due to dephasing is difficult to distinguish from a larger distribution of interaction energies than accounted for in a mean-field description. The distribution of nearest-neighbor separations in a homogenous bulk gas, for example will lead to a variation in interaction energy on the order of the average interaction energy \cite{dad15}. So for this level of modification of the spectrum, the dephasing rate $\Gamma_{Ryd}$ must be interpreted as a phenomenological parameter.}
%For this level of modification of the spectrum, the dephasing rate $\Gamma_{Ryd}$ must be interpreted as a phenomenological parameter. One possible microscopic origin for the broader spectra could be the variation of interactions due to the distribution of nearest-neighbor separations in even a homogeneous bulk gas. This variation can be on the order of the interaction strength itself \cite{desalvothesis}.
The atom-loss spectrum  changes dramatically  at later times,  collapsing to a single peak at line center. This total loss of coherence of the dressed states underlying the AT structure implies dephasing that is much greater than the coupling Rabi frequency $\Omega_{12}$ and much greater than the phenomenological dephasing rates that reproduced the short-time spectra. We cannot make a strong statement regarding the form and origin of the dephasing term that might be required to describe this data.  However, it is clear that decoherence of the Rydberg level is playing an important role, especially for longer excitation times.

%\onecolumngrid

\begin{figure*}[t]
\centering
\includegraphics[clip=true,keepaspectratio=true,width=7.2in,trim=0in 0in 0in 0in]{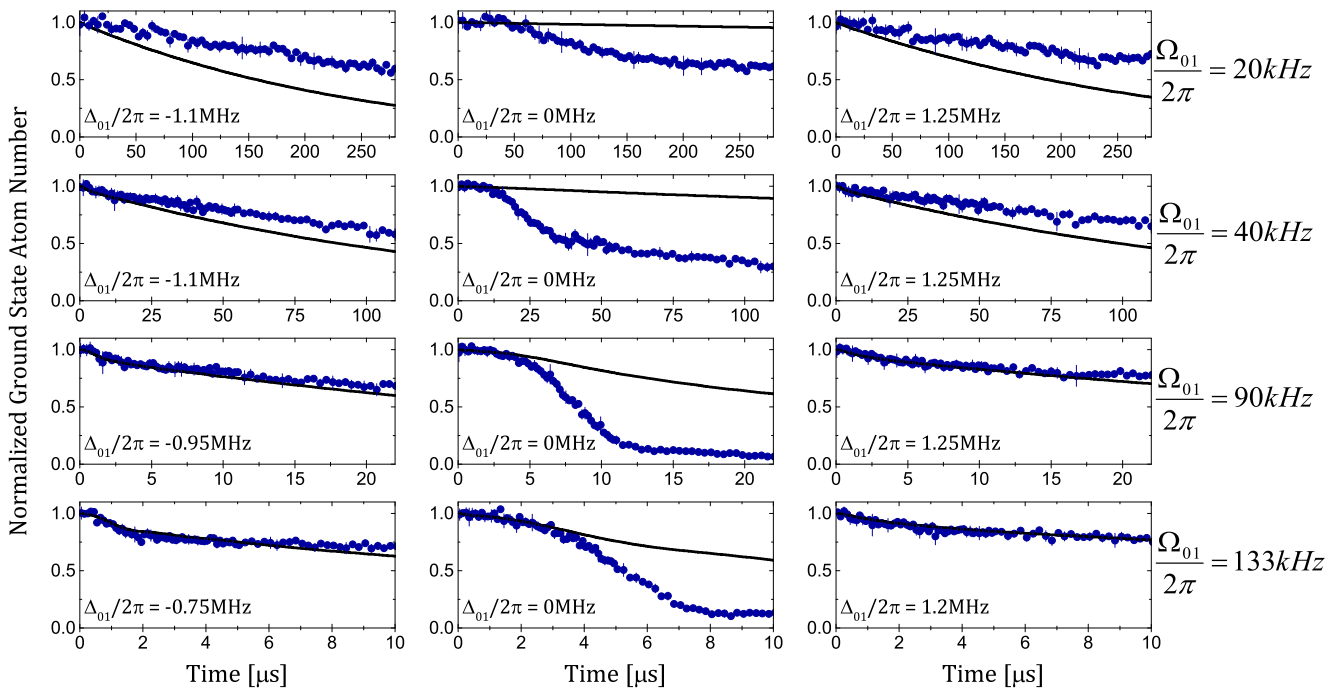}
\caption{Fractional number of atoms remaining versus laser excitation time for the 689-laser tuned to the red-detuned (Left) and blue-detuned (Right) AT peaks and for $\Delta_{01}=0$ (Center). The initial peak density is $\rho_0 = 1 \times 10^{13}$ cm$^{-3}$.  Detunings of the 689-nm laser are indicated in each figure. Blue circles are experimental data.  The lines show the results of Eq.\ \ref{eq:OpticalBloch} for $V_{Ryd}=\Gamma_{Ryd}={4\pi C_6 \rho}/{3 \hbar R_B^3}$.  }
\label{fig:HDTimeEvolution}
\end{figure*}

%\newpage
%\twocolumngrid

%\newpage
%\newpage

\subsection{Time Evolution}

%We also show the evolution in time with greater temporal resolution and longer exposure times for high density ($1 \times 10^{13}$ cm$^{-3}$) and 689-nm laser intensities corresponding to $\Omega_{01}/2\pi=\{20,56,92,133\}$\,kHz. These measurements are performed at fixed detuning corresponding to two-photon resonance ($\Delta_{01}=\Delta_{12}=0$, Fig.\ \ref{fig:HDTimeEvolution}(middle)), and
%%the red-detuned and blue-detuned loss peaks Fig.\ \ref{fig:HDTimeEvolution}(left) and (right) respectively) observed in the 2\,$\mu$s spectra.
%at the position of peak loss for the red and blue detuned Aulter-Townes peaks observed in the 2\,$\mu$s spectra (Fig.\ \ref{fig:HDTimeEvolution}(left) and (right) respectively).

The  increase in dephasing rate with time can be seen more clearly by directly measuring the time evolution of atom loss
with greater temporal resolution and longer exposure times for high density ($1 \times 10^{13}$ cm$^{-3}$) and 689-nm laser intensities corresponding to $\Omega_{01}/2\pi=\{20,40,90,133\}$\,kHz (Fig.\ \ref{fig:HDTimeEvolution}).
These measurements were performed at
  the center of the spectrum (Fig.\ \ref{fig:HDTimeEvolution}(middle)) and on resonance with the red- (Fig.\ \ref{fig:HDTimeEvolution}(left)) and blue-detuned (Fig.\ \ref{fig:HDTimeEvolution}(right)) AT peaks   observed in the 2\,$\mu$s spectra.

With $\beta=1$, we obtain good agreement between data and model at early times for all experimental conditions. This agreement extends to longer times for detunings on the AT peaks.
It is important to note that at different laser excitation frequencies the relative populations of Rydberg and low-lying states are very different, which might result in different dynamics and dominant effects. Near the center of the spectrum at $\Delta_{01} = 0$ we observe that theory substantially underestimates the loss at later times. In fact, it appears that the system is well described by moderate dephasing for a short time, which is longer for weaker excitation. Then, the system  shifts to a different behavior characterized by dramatically increased dephasing. After the overall density of atoms drops below some threshold value, which is lower for stronger excitation, the system appears to revert to the behavior characterized by less dephasing and atom loss. For example this latter transition occurs after 35\,$\mu$s of excitation for $\Omega_{12}/2\pi=56$\,kHz.

%\begin{figure}[ht]
%\centering
%\includegraphics[width = 3 in]{LowRabiLongSpectra.eps}
%\caption{Blue circles: Experimental data of ground state population after a single variable length pulse with initial peak density, $\rho_0 = 1 \times 10^{13}$ cm$^{-3}$ and $\Omega_{01} = 20.14$kHz.  Details in text.  Red lines: Theoretical calculation using LDA for varying strengths $V_{Ryd}$ and $\Gamma_{Ryd}$.  $\alpha$ and $\beta$ are defined in text.}
%\label{fig:LDTimeSpectra}
%\end{figure}

%\begin{figure}[ht]
%\centering
%\includegraphics[width = 3 in]{LDTimeEvolution.eps}
%\caption{Blue circles: Experimental data of time dependent ground state population after a single variable length pulse with initial peak density, $\rho_0 = 1.9 \times 10^{12}$ cm$^{-3}$.  Red lines: Theoretical calculation using LDA for $\alpha = 1$ and $\beta = 2$.}
%\label{fig:LDTimeEvolution}
%\end{figure}

\subsection{Possible Explanations for the Increased Dephasing}

The search for a definitive explanation for the large  observed dephasing will be a topic of a future study, but we mention a few possibilities here.
 The general feature of a delayed turn-on of a very large dephasing rate is consistent with the effects of superradiance \cite{wpa08} out of the $5s24s{^3S_1}$ Rydberg state, and/or dipole-dipole interactions \cite{avg98} between $5s24s{^3S_1}$ atoms and
atoms in nearby Rydberg $P$ states  populated by natural decay and blackbody radiation.

Other possible sources of  dephasing are DC Stark shifts and inelastic collision processes due to free charges present in the excitation volume. Free charges could arise from photoionization of Rydberg atoms, collisions of Rydberg atoms with hot background gas atoms, blackbody radiation, or Penning ionization \cite{kpp07}. The DC Stark shift for the $5s24s{^3S_1}$ Rydberg state is to the red, which would also shift the AT spectrum to the red as a function of $\Delta_{01}$. This is inconsistent with the general blue shift that is observed.
%The most rapid inelastic collision process in the system is $\ell$-changing collisions between charged particles and Rydberg atoms, which can be very rapid  \cite{dfw01}.

\begin{figure}[t]
\centering
\includegraphics[width = 3 in]{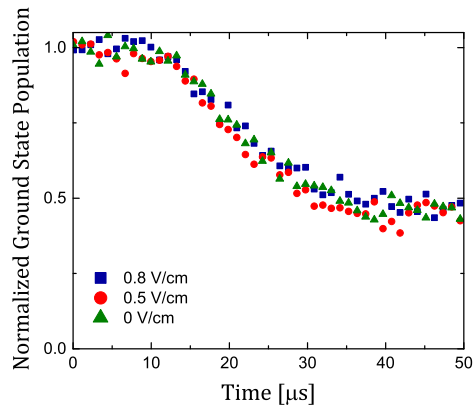}
\caption{Time evolution of atom loss at $\Delta_{01} = 0$, $\Omega_{01}/2\pi = 40$ kHz,  peak density  $\rho_0=1 \times 10^{13}\, \mathrm{cm}^{-3}$, and the indicated applied electric fields. }
\label{fig:Efield}
\end{figure}

It is beyond the scope of this paper to model collisional processes involving charged particles, but
we looked for their influence  by
 testing the effect of an electric field of  0.8\,V/cm on the time evolution of trap loss at a fixed detuning of $\Delta_{01} = 0$ and $\Omega_{01} = 40$ kHz.   For this field, an ion escapes from the trap on a timescale of 0.1 $\mu$s, which is much faster than the atom-loss timescale of tens of $\mu$s. If charged particles were important for the dynamics, the dephasing would be much smaller in the presence of a clearing field. As can be seen from the data plotted in Fig.\ \ref{fig:Efield}, however, the presence of an applied electric field yielded no effect on the data.  This strongly suggests that charged particles are not the cause of the observed dephasing.

\section{Conclusions}

We have presented an  experimental study, supported by theoretical modeling, of the effects of  Rydberg-Rydberg interactions on the AT spectrum in an ultracold gas of strontium atoms. Results  show clearly distinguishable effects associated with shifts and dephasing  of the Rydberg level that increase with density and with the Rydberg excitation fraction. We also present an effective potential for the {Rydberg level that augments mean-field theory to incorporate the effects of short-range spatial correlations  arising from the Rydberg-blockade.}
With this potential, the density dependence and excitation-strength dependence of loss spectra at short excitation times   can be  explained with a density matrix treatment. The local density approximation is used to treat the density inhomogeneity of the trapped atom sample.
At longer excitation times, the dephasing of the Rydberg level increases dramatically, especially for excitation directly on two-photon resonance $\Delta_{01}=\Delta_{12}=0$. We suggest superradiance or dipole-dipole interactions as possible explanations for the large dephasing rates.

%As a more stringent test of our theory, we will see how the spectra evolves in time.  For this data, we have used only high density samples and chose a low and high Rabi frequency to see how the spectrum changes as we increase the pulse length from 2 $\mu s$.  The results are plotted in figures \ref{fig:LDTimeSpectra} and \ref{fig:HDTimeSpectra}.

\section{Acknowledgements}
We thank  R. Mukherjee for helpful discussions on modeling Rydberg-Rydberg interactions, and we thank S, Rolston, T. Porto and E. Goldschmidt for stimulating conversations on sources of enhanced dephasing in these experiments.

Research supported by the AFOSR under grant no. FA9550-12-1-0267, the NSF under grants nos. 1301773 and 1205946, and the Robert A. Welch Foundation under grants nos. C-0734, C-1844, and C-1872, and by the FWF (Austria) under grant no P23359-N16 and by the SFB-NextLite.  The Vienna Scientific Cluster was used for the calculations.

\end{document}